\documentclass{ws-ijgmmp}

\begin{document}

\markboth{D. MOMENI at al.}
{Reciprocal NUT spacetimes}

%
\catchline{}{}{}{}{}
%

\title{ Reciprocal NUT spacetimes
}

\author{Davood Momeni\footnote{Corresponding author}}

\address{Eurasian International Center for Theoretical Physics and Department of General
Theoretical Physics, Eurasian National University, Astana 010008, Kazakhstan\\
\email{d.momeni@yahoo.com} }

\author{Surajit Chattopadhyay}

\address{Pailan College of Management and Technology, Bengal
Pailan Park, Kolkata-700 104, India\\
surajcha@iucaa.ernet.in}

\author{Ratbay Myrzakulov}

\address{ Eurasian International Center for Theoretical Physics and Department of General
Theoretical Physics, Eurasian National University, Astana 010008, Kazakhstan\\
\email{rmyrzakulov@gmail.com} }
\maketitle

\begin{history}
\received{(Day Month Year)}
\revised{(Day Month Year)}
\end{history}

\begin{abstract}
In this paper, we study the \texttt{Ehlers' transformation}  (sometimes called  gravitational duality rotation) for \texttt{reciprocal} static metrics. First we introduce the concept of reciprocal  metric. We prove a theorem which shows how we can construct a certain new static solution of Einstein field equations using a seed metric. Later we investigate the family of stationary spacetimes of such reciprocal metrics. The key here is a theorem from Ehlers', which relates any static vacuum solution to a unique stationary metric. The stationary metric has a magnetic charge. The spacetime represents Newman -Unti-Tamburino (NUT)  solutions. Since any stationary spacetime can be decomposed into a $1+3$ time-space decomposition, Einstein field equations for any stationary spacetime can be written in the form of Maxwell's equations for gravitoelectromagnetic fields. Further we show that this set of equations is invariant under  reciprocal transformations. An additional point is that the NUT charge changes the sign. As an instructive example, by starting from the reciprocal Schwarzschild as a spherically symmetric solution and reciprocal Morgan-Morgan disk model  as seed metrics we find their corresponding stationary space-times.   Starting from any static seed metric, performing the reciprocal transformation and by applying an additional Ehlers' transforation we obtain a family of NUT spaces with negative NUT factor (reciprocal NUT factors).
\end{abstract}


\section{Introduction}
Exact solutions of Einstein gravitational field equations are represented by Lorentzian manifolds with metric $g_{\mu\nu}(x^{\rho}),\mu,\nu,\rho....=0,1,2,3,..$, satisfies the common form of Einstein equation:
\begin{equation}
R_{\mu\nu}-\frac{1}{2}g_{\mu\nu}R=\frac{8 \pi G}{c^4}T_{\mu\nu}
\end{equation}
Here  $g_{\mu\nu}$, is a measure of theory . The Ricci tensor $R_{\mu\nu}=g^{\rho\sigma}R_{\rho\mu\sigma\nu}$, where $g^{\mu\rho}g_{\rho\nu}=\delta^{\mu}_{\nu}$(see  \cite{GR1} for a complete list of possible solutions). Based on different forms of energy momentum tensor, the following are the types of  well-known exact solutions: (i) Vacuum solutions, (ii) Electrovacuum solutions, (iii) Null dust solutions, (iii) Fluid solutions, (iv) Scalar field solutions and (v) Lambda-vacuum solutions \cite{GR1,GR2}. As a natural extension of the vacuum spherically symmetric solution to stationary systems, the NUT solution is given (in $t,~r,~\theta,~\phi$ coordinates) in the following form \cite{NUTfunda}
\begin{equation}
ds^2=(r^2+l^2)(d\theta^2+\sin^2\theta d\phi^2)-f(r)(dt+2l\cos\theta d\phi)^2+f^{-1}(r)dr^2.
\end{equation}
where, $f(r)=\frac{r^2-mr-l^2}{r^2+l^2}$ and $l$ defines the magnetic mass (charge) or NUT factor and $m$ is the point mass in Schwarzschild space \cite{NUTfunda}. Further  $d\hat{r}^2$ denotes the metric of unit sphere. NUT spacetime is a vacuum solution but is not asymptotically Euclidean so we can't construct a field theory on it. NUT-Space was discovered in Ehlers'’ thesis \cite{1957}and later rediscovered \cite{NUTfunda}. Diffferent aspects of this solutions have been studied in the literature\cite{NUT1,NUT2,NUT3,NUT4,NUT5}.\par
Derivation of NUT spacetime using Ehlers' transformation is known. In the present work we study a more complex version of such transformation. We start by a static seed metric which solves Einstein field equations in vacuum. We present and prove a theorem from Buchdahl \cite{a},it shows that how we can construct another static vacuum solution of Einstein gravity,which is not a formal  coordinate  transformation of the old one. The new metric is called reciprocal. We explain by reciprocal what we mean physically.  As a motivated idea
we use the Ehlers' transformation  along with the $(1+3)$ decomposition of stationary spacetimes (gravitoelectromagnetic) for reciprocal metrics . We obtained reciprocal Schwarzschild metric as a reciprocal analogue of the original one. Further, we derive reciprocal Morgan-Morgan-NUT   as seed metric of a generalized \texttt{reciprocal} Morgan-Morgan-NUT with negative NUT parameter. It should be mentioned that we shall use units in which  $i,j,a,b,..=1...3$(spatial) while $\mu,\nu,..=0..3$ and $c=G=1$. Rest of the paper is organized as follows:~In section II, starting from a static (time independent) vacuum solution of Einstein field equations, we perform a reciprocal transformation on it. In section III, we have elaborated the basic idea of Ehlers' transformation to find a way on how we can generate a certain family of stationary metrics from the given static metrics. Duality in gravitoelectromagnetism and reciprocal transformation has been discussed in section IV. In sections V, VI, VII and VIII we have discussed the reciprocal  Schwarzschild spacetime, Reciprocal NUT space from reciprocal Schwarzschild through Ehlers' transformation, Reciprocal Morgan-Morgan disk space and Reciprocal Morgan-Morgan-NUT respectively. We have concluded in section X.
\section{Reciprocal static metrics }
At the beginning of the present section we first state the main proposal of the reciprocal principle. We follow the terminology of \cite{a}. The concept of metric $\bar{g}$ reciprocal to metric $g$ was defined by the following $n\ge 4$-dimensional metric:
\begin{eqnarray}
g_{\mu\nu}dx^{\mu}dx^{\nu}=g_{ij}(x^k)dx^idx^j+g_{aa}(dx^a)^2,\  \ (\textbf{seed})\Longleftrightarrow \bar{g}=(g_{aa})^{\frac{2}{n-3}}g_{ij}dx^idx^j+\frac{(dx^a)^2}{g_{aa}},\  \ (\textbf{reciprocal}).
\end{eqnarray}
In vacuum , $R_{\mu\nu}=0$.  Both metrics $g_{\mu\nu},\bar{g}_{\mu\nu}$ must be static. It means a coordinate $x^a$ exists such that the metric $g_{\mu\nu}$ satisfies the conditions:
\begin{eqnarray}
g_{ia}=0,\ \ \frac{\partial g_{\mu\nu}}{\partial x^a}=0.
\end{eqnarray}
The reciprocal algorithm works if $g,\bar{g}$ satisfy the vacuum Einstein equation. Under reciprocal transformation the Einstein field equations are invariant. If we denote by $T_{\mu}^{\nu}=\frac{1}{2}\delta_{\mu}^{\nu}R-R_{\mu}^{\nu}$ the corresponding tensor-density,then reciprocal algorithm follows by these identities:
\begin{eqnarray}
\bar{T}_{i}^{k}=T_{i}^{k},\ \ \Big(\bar{T}_{a}^{a}-\frac{1}{n-3}\bar{T}_{i}^{i}\Big)=-\Big(T_{a}^{a}-\frac{1}{n-3}T_{i}^{i}\Big)\label{T}.
 \end{eqnarray}
If we define $Q_{\mu}^{\nu}=T_{\mu}^{\nu}-\frac{1}{n-3}\delta_{\mu}^{\nu}T_{i}^{i}$, then (\ref{T}) may be written in the form:
\begin{eqnarray}
\bar{Q}_{i}^{k}=Q_{i}^{k},\ \ \bar{Q}_{a}^{a}=-Q_{a}^{a}\label{Q}.
\end{eqnarray}
So we conclude as the following:
Reciprocal transformation of two static metrics is a transformation which preserves the form of Einstein equation and additionaly , under this transformation the components of the quantity $Q_{\mu}^{\nu}$ decomposed to $Q_{i}^{j}$ (tensor),$Q_{a}^{a}$(scalar) behave like  (\ref{Q}). If we take $x^a=t$ (time), we can underestand the scalar $Q_{a}^{a}$ as energy.\par
After this introduction to the topic we follow our search to find the four dimensional examples.
 If we choice $n=4,a=0$,we start from a static (time independent) vacuum solution of Einstein field equations in the following representation:
\begin{eqnarray}\label{rec1}
g\equiv g_{ik}dx^idx^k+g_{00}(dx^0)^2\label{g0}.
\end{eqnarray}
It solves vacuum field equations $G_{\mu\nu}(g)=0$. We have a theorem about such metrics\\
\textbf{Theorem}: Starting from (\ref{g0}) as seed metric, we perform a reciprocal transformation on it. The following new metric is also an exact solution of vacuum field equations $G_{\mu\nu}(\bar{g})=0$:
\begin{eqnarray}
\bar{g}\equiv (g_{00})^2g_{ik}dx^idx^k+\frac{(dx^0)^2}{g_{00}}\label{g1}.
\end{eqnarray}\par
\textbf{Proof}: To prove this theorem it is needed that we pass some intermediate steps by some lemma.\par
\textbf{lemma.a}: Let the general $n$ dimensional metric be in the following static form:
\begin{eqnarray}
g_{\mu\nu}dx^{\mu}dx^{\nu}=\hat{g}_{ik}dx^idx^k+e^{2\gamma}(dx^a)^2.
\end{eqnarray}
The Ricci tensor associated to this spacetime $V_{n}$ can be written in the following forms:
\begin{eqnarray}
R_{ik}=\hat{R}_{ik}+\beta_{ik},\ \ \beta_{ik}=\nabla_{i}\nabla_{k}
\gamma+\nabla_i\gamma\nabla_k\gamma\label{2.1}\\
R_{ia}=0,R_{a}^{a}=\hat{g}^{ik}\beta_{ik}=\beta.
\end{eqnarray}
Here $\hat{R}_{ik}$ is the Ricci tensor of metric $\hat{g}$. 
\par
\textbf{lemma.b}: Now if we write $\hat{g}_{ik}=e^{2\sigma}g^{0}_{ik}$ and ,$g^{0}_{ik}$ is metric of $V^{0}_{n-1}$. $V_{n-1}$ is conformal to $V^{0}_{n-1}$,so we have:
\begin{eqnarray}
\hat{R}_{ik}=R^{0}_{ik}+(n-3)(\nabla_{i}\nabla_{k}
\sigma-\nabla_i\sigma\nabla_k\sigma)+g^{0}_{ik}\Big[g^{0,lm}
\nabla_{l}\nabla_{m}
\sigma+(n-3)g^{0,lm}\nabla_{l}\sigma\nabla_{m}
\sigma\Big]\label{2.3}.
\end{eqnarray}
Here 
$$\nabla_{i}\gamma=\partial_{i}\gamma,\nabla_{i}\nabla_{k}
\gamma=\partial_{ik}\gamma-(\nabla_{i}\sigma\nabla_{k}\gamma
+\nabla_{k}\sigma\nabla_{i}\gamma)+g^{0}_{ik}(g^{0,lm}\nabla_l\sigma\nabla_m\gamma).$$
Here $g^{0,lm}g^{0}_{lk}=\delta^{m}_{k}$.\par
\textbf{lemma.c}: Using (\ref{2.3})in (\ref{2.1}) ,the Ricci tensor of $V_n$ becomes:
\begin{eqnarray}
R_{ik}=R^{0}_{ik}+(n-3)(\nabla_{i}\nabla_{k}
\sigma-\nabla_i\sigma\nabla_k\sigma)+\beta_{ik}-
(\nabla_{i}\sigma\nabla_{k}\gamma
+\nabla_{k}\sigma\nabla_{i}\gamma)\\ \nonumber+g^{0}_{ik}\Big[g^{0,lm}
\nabla_{l}\nabla_{m}
\sigma+(n-3)g^{0,lm}\nabla_{l}\sigma\nabla_{m}
\sigma+g^{0,lm}\nabla_{l}\sigma\nabla_{m}
\gamma\Big]\label{2.5}.
\end{eqnarray}
\par
Now we prove our theorem. Suppose that $R_{\mu\nu}=0$. By algebraic manipulation we arrives to the following equations:
\begin{eqnarray}
g^{0,lm}
\nabla_{l}\nabla_{m}
\mu+g^{0,lm}\nabla_{l}\mu\nabla_{m}
\mu+\frac{n-3}{4(n-2)}R^{0}=0,\\
g^{0,lm}
\nabla_{l}\nabla_{m}
\lambda+g^{0,lm}\nabla_{l}\lambda\nabla_{m}
\lambda+\frac{n-3}{4(n-2)}R^{0}=0.
\end{eqnarray}
Here $\lambda=\frac{n-3}{2}\sigma,\mu=\gamma+\frac{n-3}{2}\sigma$. Using these expressions,we are able to rewrite (\ref{2.5}) in the following equivalent form:
\begin{eqnarray}
R_{ik}=R^{0}_{ik}+\nabla_{i}\nabla_{k}\lambda+\nabla_{i}\nabla_{k}
\mu+
\nabla_{i}\lambda\nabla_{k}\lambda-\frac{n-1}{n-3}\Big(\nabla_{i}\lambda\nabla_{k}\mu+\nabla_{i}
\mu\nabla_{k}\lambda\Big)+\nabla_{i}\mu\nabla_{k}\mu\\ \nonumber+2g^{0}_{ik}
\Big(\frac{1}{n-3}g^{0,lm}\nabla_{l}\lambda\nabla_{m}\mu
-\frac{1}{4(n-2)}R^{0}\Big)\label{3.3}.
\end{eqnarray}
The above equation shows that the Einstein equations $R_{\mu\nu}=0$ are invariant with respect to the mutual interchange of $\lambda,\mu$. \par
We note that (\ref{3.3}) satisfies by two set of metric functions:
\begin{eqnarray}
\lambda=0,\ \ \mu=\frac{1}{2}\log g_{aa}.
\end{eqnarray}
This is the seed metric. The next solutions is:
\begin{eqnarray}
\sigma=\frac{1}{n-3}\log g_{aa},\ \ \gamma=-\frac{1}{2}\log g_{aa}.
\end{eqnarray}
This is the reciprocal metric $\bar{g}$. It completes our proof.
\par
Transformation between two kinds of metrics given by Eqs.(\ref{g0}) and (\ref{g1}) is called reciprocal transformation. It was proposed firstly for vacuum solutions \cite{a}. However, later it was extended to non-vacuum cases also. In the present work we shall investigate the vacuum solutions through reciprocal transformation in the simplest form as shown in the above theorem\cite{a}.

\section{Ehlers' transformation}
The basic idea behind Ehlers' transformation is to find a way for generating a certain family of stationary metrics from the given static (non-stationary) metrics. Let us assume that we have a static mass distribution density with gravitational dipole moments. If this mass configuration is rotated physically around an axis then it transits to higher order moments, which are quadrapoles in this case. We are interested to understand this kind of transition in the language of  electro-magnetic gravitational fields. It means we find a similar formulation of the Einstein field equations in vacuum in terms of a set of the vector fields, in a similar manner as electric or magnetic fields in classical Maxwell's theory. Indeed, as we shall observe, transition of a dipole moment to quadrapole is well understood in terms of the gravitoelectromagnetic fields. To start, let us introduce a methodology using
the $(1+3)$-decomposition (threading) of a spacetime,proposed generally for any stationary spacetime  via a
congruence family of time like curves.
For a typical metric we obtain
following decomposion  for the spacetime line element \cite{Landau} :
$$ds^{2}=dT^{2}-dL^{2},\eqno(1)$$
In the terminology of \cite{Landau}, here $dL$ and $dT$ are defined to be {\it the invariant spatial
and temporal length elements} under general coordinate transformation. It defines the spacetime line element between two  causal  events.
They are obtained from a set of normalized tangent vector $T^{a}=
{\xi^a \over |\xi|}$  to the time like curves in the following form
;
$$dL^{2}= h_{ab}dx^{a}dx^{b},\eqno(2)$$
$$dT= u_{a}dx^{a},\eqno(3)$$
Here $h_{ab}$ denotes an induced metric
$$h_{ab}= -g_{ab}+u_{a}u_{b},\eqno(4)$$
Here we define a vector potential using  $A_a  \equiv -{\xi_a \over |\xi|^2} $. We are able to rewrite the metric in the following optional form:
$$ds^{2}=  h(A_a {\rm d}x^a)^2 - h_{ab}{\rm d}x^a
{\rm d}x^b\;\;\; ; \;\;\; h_{ab}= -g_{ab}+h A_{a}A_{b},\ \ h=|h_{ab}|.\eqno(5)$$
We opt for a well posed
and {\it preferred parameterization of the coordinate time} $x^{0}$ (generally and strictly speaking it doesn't have the same interpretation as physical time) for the comoving
observers;
$$\xi^a = (1,0,0,0)\;\;\;\;\;\ ; \;\;\;\;
A_a = (-1, - {g_{0\alpha}\over g_{00}}),\eqno(6)$$
 So far, we have the following
equations\cite{De-Felice,Landau};
$$dL^2 \equiv dl^{2}=\gamma_{ab}dx^{a}dx^{b},\eqno(7)$$
$$ds^{2}=e^{2U}(dx^{0}-A_{a}dx^{a})^{2}-dl^{2},\eqno(8)$$
where
$$e^{2U}\equiv g_{00} \;\;\;\;\; ,\;\;\;\;\; A_{a} = -g_{0a}/g_{00},\eqno(9)$$
and
$$\gamma_{ab} = (-g_{ab} +{g_{0a} g_{0b}\over g_{00}}).\eqno(10)$$
We introduce $U$ and $A$ as the {\it gravitoelectric}
and {\it gravitomagnetic} potential functions respectively. The corresponding vector fields are\cite{RMP};
\begin{eqnarray}
{{\bf E}_g} = -\nabla {U}\label{E},
\\{\bf B}_g = {\rm curl} {\bf A }\label{B}.
\end{eqnarray}
where the vector differential operators are computed in the $\gamma$ space. Einstein gravitational field equation in vacuum has the following forms:
\begin{eqnarray}
\nabla.\bf{B}_g=0,\\
\nabla\times \bf{E}_g=0,\\
\nabla.{\bf{E}}_g=e^{2U} \frac{{\bf{B}}_g^2}{2}+\bf{E}_g^2,\\
\nabla\times (e^{U}{ \bf{B}_g })=2e^{U} {\bf{E}}_g\times {\bf{B}}_g.
\end{eqnarray}
We have the following theorem:\par
\textbf{Ehlers' Theorem}\cite{Ehlers'}:
If

\begin{eqnarray}
g_{\mu\nu}dx^{\mu} dx^{\nu}= e^{2U}(dx^0)^2 -e^{-2U}d{\tilde l}^2\label{seed}.
\end{eqnarray}

with $d{\tilde l}^2 = e^{2U} dl^2$  representing the {\it conformal} spatial distance,
represents the metric of a static vacuum spacetime, then,
\begin{eqnarray}
\bar{g}_{\mu\nu}dx^{\mu} dx^{\nu} = \frac{1}{\alpha {\rm cosh}(2U)} (dx^0 - A_a dx^a)^2
- \alpha {\rm cosh}(2U)d{\tilde l}^2\label{NUT}.
\end{eqnarray}

(with $\alpha = constant > 0$, $U=U(x^a)$ and $A_b =
A_b(x^a)$) is the metric of a stationary vacuum spacetime
{\it if and only if  } $A_a$ satisfies the following equation;
\begin{eqnarray}
\alpha\sqrt{\tilde \gamma} \varepsilon_{abc} U^{,c} =
A_{[a,b]}\label{Ehlers'-eq}.
\end{eqnarray}
where $\tilde \gamma = {\rm det}{\tilde \gamma}_{ab}$, ${\tilde \gamma}_{ab}$
is the conformal spatial metric.
Eq. (\ref{Ehlers'-eq}) is called as the Ehlers' equation. Integrationg of Ehlers' equation is equivalent to solve Maxwell's  equation by a given scalar potential $U$.
In this work,  we shall derive the reciprocal NUT
spaces as the stationary spacetime
which is the gravitational dual of reciprocal static space as seed.
\section{Properties of  gravitoelectromagnetism equations under reciprocal transformation}
Throughout this section we used $\hat{}$ to refer to the reciprocal quantities .
By simple replacement of $\hat{U}= -U$ the reciprocal form of (\ref{seed}) is obtained as:
\begin{eqnarray}
g_{\mu\nu}dx^{\mu} dx^{\nu}= e^{-2U}(dx^0)^2 -e^{2U}d{\tilde l}^2\label{rec-seed}.
\end{eqnarray}
One can write (\ref{NUT}) analogous to (\ref{rec-seed}) in the form:
\begin{eqnarray}
{\bar g}_{\mu\nu}dx^{\mu} dx^{\nu} = \frac{1}{\alpha {\rm cosh}(2U)} (dx^0 - \hat{\bf{A}}_a dx^a)^2
- \alpha {\rm cosh}(2U)d{\hat l}^{2} \label{rec-NUT}.
\end{eqnarray}
The reciprocal vector field $\hat{A}$ satisfies the reciprocal Ehlers' equation:
\begin{eqnarray}
-\alpha\sqrt{\hat{\gamma}}\varepsilon_{abc} U^{,c} =
\hat{\bf{A}}_{[a,b]}\label{rec-Ehlers'-eq}.
\end{eqnarray}
Note that $d\hat{\gamma}^2=g_{00}g_{ij}dx^idx^j=d{\tilde l}^2$,consequently $\sqrt{\tilde{\gamma}}=\sqrt{\hat{\gamma}}$. Comparison of (\ref{Ehlers'-eq},\ref{rec-Ehlers'-eq}) leads to:
\begin{eqnarray}
\hat{\bf{A}}=-\bf{A}.
\end{eqnarray}
So, gravitational dual of the reciprocal metric changes the gravito fields to the following forms:
\begin{eqnarray}
{\hat{{\bf E}}_g}=-\bf{E}_g\label{rec-E}\\
\hat{\bf{B}}_g=-\bf{B}_g\label{rec-B}.
\end{eqnarray}
The reciprocal forms of the field equations are written in the following:
\begin{eqnarray}
\nabla.\hat{\bf{B}}_g=0,\\
\nabla\times \hat{\bf{E}}_g=0,\\
\nabla.\hat{\bf{E}}_g=e^{2U}\frac{\hat{\bf{B}}_g^2}{2}+\hat{\bf{E}}_g^2,\\
\nabla\times (e^{U}\hat{\bf{B}}_g )=2e^{U}\hat{\bf{E}}_g\times \hat{\bf{B}}_g.
\end{eqnarray}
Note that $e^{2U}|_{1+3}=\frac{1}{\alpha\cosh(2U)}|_{Ehlers'}$. Under reciprocal transformations the system of the field equations remain invariant if and only if $\hat{\alpha}=-\alpha$. 
To preserve the same forms of the equations for $(\hat{\bf{E}},\hat{\bf{B}})$ under reciprocal transformation, we must have:
\begin{eqnarray}
e^{2\hat{U}}|_{1+3}=-e^{2U}|_{1+3},\hat{U}=-U\Rightarrow \frac{1}{\hat{\alpha}\cosh(2\hat{U})}|_{Ehlers'}=-\frac{1}{\alpha\cosh(2U)}|_{Ehlers'}\Rightarrow \hat{\alpha}=-\alpha.
\end{eqnarray}
\par
{\bf{Theorem}}:
The full system of the Maxwell equations for gravitoelectromagnetism fields is invariant under a reciprocal transformation of the fields and NUT parameter as the follows:
\begin{eqnarray}
\hat{U}=-U\\
{\hat{{\bf E}}_g}=-\bf{E}_g\label{rec-E}\\
\hat{\bf{B}}_g=-\bf{B}_g\label{rec-B},\\
\hat{\alpha}=-\alpha.
\end{eqnarray}
 So the original seed metric and reciprocal one, are physically correspond to two different spacetimes .

\section{Reciprocal  Schwarzschild spacetime}
Reciprocal metric of Schwarzschild spacetime is obtained using the reciprocal formalism. One has:
\begin{eqnarray}
\bar{g}=(1-\frac{2M}{r})^{-1}(dx^{0})^2-(1-\frac{2M}{r})dr^2-r^2(1-\frac{2M}{r})^2d\Omega^2 \label{schwz-rec}.
\end{eqnarray}
This is verified to be a non flat  exact solution to the vacuum Einstein field equations using the Maple tensor package.
\footnote{{\it Schwarzschild is unique solution which could be cast in the isotropic coordinates.
Addition of a constant in potential in Einstein gravity is non-trivial, Schwarzshild solution is $g_{00} = -1/g_{11} = 1 + 2\Phi$ where $\Phi = -M/r$. If one adds a constant to $\Phi$ to write $\Phi=-M/r + k$, then it would no longer be a vacuum solution}}.

 We mention here that the coordinate transformation $r=-\bar{r}+2M$ transforms (\ref{schwz-rec}) to the common form of the Schwarzschild spacetime:
\begin{eqnarray}
\bar{g}=(1-\frac{2M}{\bar{r}})(dx^0)^2-(1-\frac{2M}{\bar{r}})(d\bar{r})^2-(\bar{r})^2d\Omega^2.
\end{eqnarray}

\section{Reciprocal NUT space from reciprocal Schwarzschild through Ehlers' transformation}
We begin with the reciprocal Schwarzschild mertic in the form (\ref{schwz-rec}) ;
$$U=-\frac{1}{2}\log(1-\frac{2M}{r}) $$
and
$$d{\tilde l}^2 =dr^2+r^2(1-\frac{2M}{r}) d\Omega^2$$.

\par

It is observed that starting by this seed metric, the Ehlers' equation (\ref{Ehlers'-eq}) reduces to,
$$2\alpha r^2\sin\theta e^{-2U}\tilde{\gamma}^{rr}U_{,r}=A_{\phi,\theta} ,\eqno(18)$$
in which we suppose that  $A_{\theta,\phi}=0$ to keep the terminated stationary spacetime single valued
and axially symmetric.\\
 By imposing axial symmetry,  gravitomagnetic field is obtained to be,
$$A_\phi(\theta) =-2M\alpha  \cos\theta+c_0$$ (it is compared with the case of non reciprocal NUT space with a plus sign \cite{Momeni:2005uc}).\par
So the resulted stationary metric is given by;
\begin{eqnarray}
ds^2=-\frac{1}{\alpha}\frac{r^2-2Mr}{r^2-Mr+2M^2}\Big(dt-(-2\alpha M \cos\theta+c_0)d\phi\Big)^2
+\alpha\frac{r^2-Mr+2M^2}{r^2-2Mr}dr^2\\ \nonumber+\alpha(r^2-Mr+2M^2)(d\theta^2+\sin\theta^2d\phi^2).
\end{eqnarray}
It coincides with the NUT space in its original form if we apply a suitable coordinate transformation \cite{Momeni:2005uc}. Indeed, this solution matches to the  previously solution was obtained in \cite{Geroch}.
The case of pure NUT is obtained when we set $M=0$ but we kept $\hat{q}=\alpha M\neq0,c_0=2\alpha M$, it reads as the following form:
\begin{eqnarray}
ds^2=-\frac{1}{\alpha}\Big(dt-2\hat{q}(1+ \cos\theta)d\phi\Big)^2
+\alpha dr^2+\alpha r^2(d\theta^2+\sin^2\theta d\phi^2).
\end{eqnarray}
If we rescale the radial coordinate as $\sqrt{\alpha}r=\tilde{r}$ and meanwhile $\tilde{t}=\frac{t}{\sqrt{\alpha}},q'=\frac{\hat{q}}{\sqrt{\alpha}}$ we obtain:
\begin{eqnarray}
ds^2=-\Big(d\tilde{t}-2q'(1+ \cos\theta)d\phi\Big)^2
+d\tilde{r}^2+ \tilde{r}^2(d\theta^2+\sin^2\theta d\phi^2).
\end{eqnarray}
Reciprocal transformation here refers to the change of the sign of potential function $U=\log\sqrt{g_{00}}$. It means the reciprocal configurations has the same amount of energy but with negative sign of magnetic charge. After applying the dual transformation, the same stationary metric is obtained but with a negative charge sign. Both stationary metrics mimic the same form but their seeds are reciprocal. Dual gravitational rotation included a NUT charge which has the opposite sign for both seed metrics, reciprocal and non reciprocal. Physically, Ehlers' transformation induced a reciprocal (with opposite sign) of gravitomagnetic monopople charge . Reciprocality in the energy content of the seed metric is related to the charge of the resulted stationary solutions.

\section{Reciprocal Morgan-Morgan disk space }
The thin metric of a static disk with gravitational Quadrapoles was introduced originally in \cite{Rosen}. A modern formulation of this metric was presented later by Morgan-Morgan (MM) spacetime \cite{MM}. The appropriate coordinate system is oblate ellipsoidal coordinates $(\xi , \eta)$
in which the definition of these coordinates  is understood through the
The Weyl $(\rho, z)$ axial coordinates are;
$$ \rho^2 = a^2(1+\xi^2)(1-\eta^2)\;\;\; , \;\;\; z=a\xi\eta \; , \;
|\eta|\leq 1 \;\;\; , \;\;\; 0\leq\xi\leq \infty, \eqno(37)$$
The location of the disk is at $\xi =0 \; , \;|\eta|\leq 1 $ .\par
 Reciprocal MM spacetime is written in the following form:
\begin{eqnarray}
\bar{g}=e^{-2u}(dx^0)^2-e^{2u}\Big[e^{2k}(d\rho^2+dz^2)+\rho^2d\varphi^2\Big]\label{MM-rec}.
\end{eqnarray}
in terms of the oblate ellipsoidal coordinates $(\xi , \eta)$;
$$u=-{M\over a}\left( arccot \xi + {1\over 4}[(3\xi^2+1)arccot \xi-3\xi]
(3\eta^2-1)\right),\eqno(34)$$
and
$$k={9\over 4}M^2\rho^2 a^{-4}\left[ ({\rho \over a})^2 B^2(\xi) -
(1+\eta^2)A^2(\xi) - 2\xi(1-\eta^2)A(\xi) B(\xi)\right],\eqno(35)$$
Here auxiliary functions are defined by
$$A(\xi) = \xi {\rm arccot} \xi -1 \;\;\; , \;\;\; B(\xi) = {1\over 2}[{\xi \over
1+\xi^2} - {\rm arccot}\xi],\eqno(36)$$
MM metric is one of the most explicit forms of the disk metrics. It has an interesting interpretation as a dust disk rotationg clock side and counter clock side in such a way that the metric has zero net angular momentum.

\section{Reciprocal Morgan-Morgan-NUT  }
Starting with the reciprocal Morgan-Morgan static disk space given by equation (\ref{MM-rec})  we firstly rewrite it as
follows
$$U=u={M\over a}\left( {\rm arccot} \xi + {1\over 2}[(3\xi^2+1){\rm arccot} \xi-3\xi]
P_2(\eta)\right).\eqno(40)$$
Now selecting an axially symmetric gravitomagnetic potential ${\bf
A}=A_\phi(\xi,\eta){\hat{\bf\phi}}$,via Ehlers' transformation (\ref{Ehlers'-eq}) we obtain;
$$A_{\phi , \eta} = 2\alpha a (1+\xi^2)U_{,\xi},\eqno(41)$$
$$A_{\phi , \xi} = -2\alpha a (1-\eta^2)U_{,\eta},\eqno(42)$$
where $\alpha$ is the duality
rotation parameter and $a$ denotes the radius of the thin disk .
By substituting the above potential into the equations we find:
\begin{eqnarray}
&&A_\phi(\xi,\eta) = 3l\xi \eta(1-\eta^2)(1+\xi^2)\arctan \xi\\ &&\nonumber+ {3\over 2} \eta l \left[(\pi \xi^3 -2\xi^2 -4/3 + \pi\xi)\eta^2  -
 (\pi \xi^2 -2\xi + \pi)\xi \right]+\beta .
\end{eqnarray}
The NUT factor for reciprocal metric changed the sign from plus to minus  in the form $l=-M\alpha$ (with the negative sign in comparison to the usual NUT space). To recover the reciprocal Morgan-Morgan space
as $l\rightarrow 0$, we adapted  $\alpha = 2\beta$,
Consequently the metric of the reciprocal stationary spacetime of a thin
disk with mass $M$ and magnetic mass (NUT factor) $q$ is obtained by :
\begin{eqnarray}
&&ds^2 = -(\frac{q^2}{4m^2}e^{-2U}+e^{2U})^{-1}\left[ dt - A_\phi (\xi,\eta)
d \phi \right]^2  \\&&
   \nonumber+
a^2(\frac{l^2}{4m^2}e^{-2U}+e^{2U})\Big(e^{2k} (\xi^2 +
\eta^2)(\frac{d\xi^2}{ 1+\xi^2} + \frac{d\eta^2}{ 1-\eta^2}) +\\&& \nonumber
(1+\xi^2)(1-\eta^2)d\phi^2\Big).
\end{eqnarray}
In a similar manner as the previous case and using the Maple tensor package, it is easy to  verify that  this metric solves vacuum Einstein field equations  .
\section{Reciprocal flat spacetimes}
As an illustrative problem, consider the following parametrization of the flat spacetime:
\begin{eqnarray}
g_{\mu\nu}dx^{\mu}dx^{\nu}=x^2dt^2-(dx^2+dy^2+dz^2).
\end{eqnarray}
The reciprocal metric for this flat space is:
\begin{eqnarray}
\bar{g}_{\mu\nu}dx^{\mu}dx^{\nu}=x^{-2}dt^2-x^4(dx^2+dy^2+dz^2).
\end{eqnarray}
This metric is non flat. 
So, reciprocal transformation gives us a curved spacetime,by starting from the flat spacetime\cite{a}.

\section{Conclusion}
Generating methods to find new exact solutions of Einstein equations are few. As a motivated work, we reintroduced and proved a theorem. This theorem stated that for any given static solution of Einstein field equations there exists a unique reciprocal metric which also satisfies field equations. The physical meaning of reciprocal transformation has been understood via energy tensor. Under reciprocal transformation the spatial components of the seed and the reciprocal metric are invariant. But the time component or energy density changed the sign. It can be interpreted as the negative mass parameter of reciprocal metric in comparison to the original one. Another important transformation is Ehlers' transformation, was called as gravitational dual transformation. It was proposed as a tool to find a corresponding stationary metric from a given static solution. The main idea is inspired by the concept of rewriting the Einstein equations in stationary spacetimes in terms of a pair of electro-magnetic fields which have gravitational origin. Following this formalism the field equations can be cast in the form of Maxwell's equations.
In the present paper we have come up to an work where, by using the Ehlers' (dual gravitational) transformation ,by starting from the reciprocal Schwarzschild  , reciprocal Morgan-Morgan-NUT  as seed metrics we find their corresponding stationary space-times. The stationary space-times obtained in this work is found to endure from a \texttt{NUT}-type family of solutions. Under reciprocal transformations , NUT  parameter changes the sign due to the fact that reciprocal seed and the original one has reciprocal mass parameters.  It has been shown that the system  of field equations for gravitoelectromagnetic fields is invariant under reciprocal transformations. An additional observation is that the NUT charge is changed the sign. Starting from any static seed metric, performing the reciprocal transformation and by applying an additional Ehlers' transforation we have obtained a family of NUT spaces with negative NUT factor (reciprocal NUT factors). We mention here that finding the reciprocal transformation for a stationary space time in the following form :
\begin{eqnarray}
g_{\mu\nu}dx^{\mu}dx^{\nu}=-N^2(x^{\mu})(dx^0)^2+g^{ab}(dx^0-N^adx_a)(dx^0+N^bdx_b).
\end{eqnarray}
Is an open problem which is under study. 

\section*{Acknowledgments}
We would like to thank the anonymous reviewer for enlightening comments related to this work.

\appendix

\end{document}